\documentclass[aps,pre,groupedaddress,showpacs]{revtex4}
\usepackage{amsmath} 
\usepackage{amsfonts}
\usepackage{amstext}
\usepackage{latexsym, amssymb} 
\usepackage{bm} %per il grassetto
\usepackage[linktocpage=true,colorlinks=true,citecolor=blue]{hyperref} %funzia con sopra il dvips
\begin{document}
\preprint{IFUM-808-FT}
\title{On the quantum description of Einstein's Brownian motion}
\author{Francesco \surname{Petruccione}}
\email{petruccione@ukzn.ac.za}
\affiliation{School of Pure and Applied Physics, Howard College,
  University of KwaZulu-Natal, Durban, 4041, South Africa}
\author{Bassano \surname{Vacchini}}
\email{bassano.vacchini@mi.infn.it}
\affiliation{Dipartimento di Fisica
dell'Universit\`a di Milano and INFN,
Sezione di Milano
\\
Via Celoria 16, I--20133, Milan, Italy}
\date{\today}
\begin{abstract}
A fully quantum treatment of Einstein's Brownian motion is given,
showing in particular the role played by the two original requirements
of translational invariance and connection between dynamics of the
Brownian particle and atomic nature of the medium. The former
leads to a clearcut relationship with Holevo's result on
translation-covariant quantum-dynamical semigroups, the latter to a
formulation of the fluctuation-dissipation theorem in terms of the
dynamic structure factor, a two-point correlation function introduced
in seminal work by van Hove, directly related to density
fluctuations in the medium and therefore to its atomistic, discrete nature. A
microphysical expression for the generally temperature dependent
friction coefficient is given in terms of the dynamic structure factor
and of the interaction potential describing the single collisions. A
comparison with the Caldeira Leggett model is drawn, especially in
view  of the requirement of translational invariance, further
characterizing general structures of reduced dynamics arising in the
presence of symmetry under translations.
\end{abstract}
%\keywords{}
\maketitle

\section{Introduction}
\label{sec:introduction}
Right a century has passed since Albert Einstein published the first
of a series of papers on the theory of Brownian
movement~\cite{Einstein,EinsteinDOVER}, a pioneering work attempting
to provide a suitable theoretical framework for the description of a
long-standing experimental puzzle~\cite{Brown}. Einstein's
investigation has gone much beyond the explanation of an interesting
experiment, proving a milestone in the understanding of statistical
mechanics of non-equilibrium processes, motivating and inspiring
physical and mathematical research on stochastic processes. By now the
term Brownian motion is ubiquitously found in the physical literature,
both at quantum and classical level, used as a kind of keyword in a
wealth of situations relying on a description in terms of mathematical
structures or physical concepts akin to those first appeared in the
explanation of Einstein's Brownian motion. In this paper we address
the question of a proper quantum description of Brownian motion in the
sense of Einstein, i.e., the motion of a massive test particle in a
homogeneous fluid made up of much lighter particles. In doing so we
actually go back to Einstein's real motivation in facing Brownian
motion, i.e., to demonstrate the molecular, discrete nature of matter.
His aim was in fact to give a decisive argument probing the
correctness of the molecular-kinetic conception of heat, a question he
considered most important, as stressed in the very last sentence of
the paper, actually quite emphatic in the original German version:
\textit{M\"oge es bald einem Forscher gelingen, die hier aufgeworfene,
  f\"ur die Theorie der W\"arme wichtige Frage zu entscheiden!}~\cite{citation}
\par
In contrast with previous approaches and results, based either on a
modelling of the environment aiming at exact solubility given a
certain suitable phenomenological Ansatz~\cite{CLPhysicaA83}, or on an
axiomatic approach relying on mathematical
input~\cite{LindbladQBM,Sandulescu}, or on the exploitation of
semiclassical correspondence~\cite{DiosiEL95}, we will base our
microscopic analysis on the two key features of Einstein's Brownian
motion: homogeneity of the background medium, reflected into the
property of translational invariance, and the atomic nature of matter
responsible for density fluctuations, showing up in a suitable
formulation of the fluctuation-dissipation relationship. Translational
invariance comes about because of the homogeneity of the fluid and the
translational invariance of the interaction potential between test
particle and elementary constituents of the fluid.
%This
%fundamental symmetry property leads to important restrictions on the
%expression for possible interactions, and given that the dynamics can
%be fairly assumed to be Markovian, on the structure of the completely
%positive generator of a quantum-dynamical semigroup describing the
%reduced dynamics of the test particle, according to recent, most
%relevant results by Holevo on the characterization of
%quantum-dynamical semigroups covariant under a suitable symmetry group,
%in this case $\mathbb{R}$, i.e., translations~\cite{HolevoJMP}. 
This fundamental symmetry property leads to important restrictions
both on the expression for possible interactions and on the structure
of the completely positive generator of a quantum-dynamical semigroup
describing the Markovian reduced dynamics.  The first key point is
therefore to consider the proper type of translational invariance
interaction leading to Einstein's Brownian motion, thus fixing the
relevant correlation function appearing in the structure of the
generator of the quantum-dynamical semigroup, which turns out to be
the so-called dynamic structure factor and provides the natural
formulation of the fluctuation-dissipation relationship for the case of
interest, first put forward by van Hove in an epochal
paper~\cite{vanHove}.  Given that the dynamics can be fairly assumed
to be Markovian, the second key point is the characterization of the
structure of generators of quantum-dynamical semigroups covariant
under a suitable symmetry group, in this case $\mathbb{R}$, i.e.,
translations, which has been recently given in most relevant work by
Holevo~\cite{HolevoJMP}.

\par
The present paper partially builds on previous
work~\cite{art3,art4,art5}, putting it in a wider conceptual and
theoretical framework, providing the previously unexplored
connection to the fluctuation-dissipation theorem and further
comparing this kinetic approach to quantum dissipation with the one by
Caldeira Leggett, also in view of recent criticism on the realm of
validity of the last approach~\cite{SolsPhysicaA94,AlickiOSID04}. In
this way a new, different approach to the quantum description of
decoherence and dissipation is put forward, which, though obviously
not universally valid, could provide a direct connection between a
precise microphysical model and reduced dynamics for a wide class of
open quantum systems, characterized by suitable symmetries. While
universality might often be a fancy, loose word in such a complex
framework, this precise microphysical modelling makes a close,
quantitative comparison between
present~\cite{ZeilingerQBM-exp,ZeilingerQBM-th1,garda03,ZeilingerQBM-th2} and next
generation experiments on decoherence and dissipation in principle
feasible.
\par
The paper is organized as follows: in Sect.~\ref{sec:transl-invar} we
introduce the basic possible translationally invariant interactions,
putting into evidence their effect on the structure of the reduced
dynamics, also in comparison with previous models in the literature;
in Sect.~\ref{sec:fluct-diss-theor} we point out the relevant
interaction for the description of Einstein's quantum Brownian motion,
showing the related expression of the fluctuation-dissipation theorem;
in Sect.~\ref{sec:quant-descr-einst} we come to the formulation of
Einstein's quantum Brownian motion putting into evidence the general
microphysical expression for the friction coefficient in terms of a
suitable autocorrelation function; in
Sect.~\ref{sec:conclusions-outlook} we finally comment on our results
and discuss possible future developments.
\section{Translational Invariance}
\label{sec:transl-invar}
As a first step we characterize the general structure of microscopic
Hamiltonians leading to a translationally invariant reduced dynamics
for the test particle. Due to translational invariance the test
particle has to be free apart form the interaction with the fluid,
subject at most to a potential linearly depending on position, e.g. a
constant gravitational field, so that in particular it has a
continuous spectrum. The fluid is supposed to be stationary and
homogeneous, and for simplicity, without loss of generality,
possessing inversion symmetry, so that energy, momentum and parity are
constants of motion.
\subsection{Characterization of translationally invariant
  interactions}
\label{sec:char-transl-invar}
The microscopic Hamiltonian may be written in the form
\begin{equation}
   \label{eq:1}
     H_{{\rm \scriptscriptstyle PM}}=H_{{\rm \scriptscriptstyle
      P}}+H_{{\rm \scriptscriptstyle
      M}}+V_{{\rm \scriptscriptstyle PM}}, 
\end{equation}
where the subscripts P and M stand for particle and matter
respectively, while $H_{{\rm \scriptscriptstyle P}}$ and $H_{{\rm
    \scriptscriptstyle M}}$ satisfy the aforementioned constraints.
The key point is the characterization of a suitable translationally
invariant interaction potential, which we put forward in the formalism
of second quantization. This non-relativistic field theoretical
approach is the natural one in order to account for statistics and
more generally many-particle features of the background macroscopic
system, also proving useful in microphysical calculations~\cite{art6}
and allowing to deal not only with the one-particle sector of
the Fock-space in which the fields referring to the test particle are
described. The interaction potential between test particle and matter
will have the general form
\begin{equation}
   \label{eq:2}
   V_{{\rm \scriptscriptstyle PM}}=
\int d^3 \! \bm{x} \int d^3 \! \bm{y} \,
A_{{\rm \scriptscriptstyle P}} (\bm{x})
t (\bm{x} - \bm{y})
A_{{\rm \scriptscriptstyle M}} (\bm{y}) ,
\end{equation}
where $t (\bm{x})$ is a $\mathbb{C}$-number, in the following
applications short range, interaction potential; $A_{{\rm
    \scriptscriptstyle P}} (\bm{x})$ is a self-adjoint operator built
in terms of the field
\begin{equation}
   \label{eq:3}
   \varphi (\bm{x})=  \int \frac{d^3 \! \bm{p}}{(2\pi\hbar)^{3/2}} \,
e^{\frac{i}{\hbar}\bm{p}\cdot\bm{x}} a_{\bm{p}}
\end{equation}
satisfying canonical commutation or anticommutation relations,
according to the spin of the test particle; similarly $A_{{\rm
    \scriptscriptstyle M}} (\bm{y})$ a self-adjoint operator given by
a  function of the field
\begin{equation}
   \label{eq:4}
   \psi (\bm{y})= \int \frac{d^3 \! \bm{\eta}}{(2\pi\hbar)^{3/2}} \,
e^{\frac{i}{\hbar}\bm{\eta}\cdot\bm{y}} b_{\bm{\eta}}
\end{equation}
pertaining to the macroscopic system and obeying suitable commutation
or anticommutation relations. Eq.\eqref{eq:2} can be most meaningfully
rewritten in terms of the Fourier transform of the interaction
potential
\begin{equation}
   \label{eq:5}
   \tilde{t} (\bm{q})=\int \frac{d^3 \! \bm{x}}{(2\pi\hbar)^{3}} \,
        e^{{i\over\hbar}\bm{q}\cdot \bm{x}}  
        t (\bm{x})
        ,  
\end{equation}
where the continuous parameter $\bm{q}$, to be seen as a momentum
transfer, has a natural group theoretical meaning as label of the
irreducible unitary representations of the group of translations, as
to be stressed later on, thus coming to the equivalent expression
\begin{equation}
   \label{eq:6}
   V_{{\rm \scriptscriptstyle PM}}=
\int d^3 \! \bm{q} \,
\tilde{t} (\bm{q})
A_{{\rm \scriptscriptstyle P}} (\bm{q})
A_{{\rm \scriptscriptstyle M}}^{\scriptscriptstyle \dagger} (\bm{q}),
\end{equation}
where the operators $A_{{\rm \scriptscriptstyle P}} (\bm{q})$ and
$A_{{\rm \scriptscriptstyle M}} (\bm{q})$ are defined according to
\begin{equation}
   \label{eq:7}
   A_{{\rm \scriptscriptstyle P/M}} (\bm{q})=\int d^3 \! \bm{x} \,
e^{-\frac{i}{\hbar}\bm{q}\cdot\bm{x}} A_{{\rm \scriptscriptstyle P/M}} (\bm{x}),
\end{equation}
so that in particular because of the self-adjointness of $A_{{\rm
    \scriptscriptstyle P/M}} (\bm{x})$ one has the identity
\begin{equation}
   \label{eq:8}
   A_{{\rm \scriptscriptstyle P/M}}^{\scriptscriptstyle \dagger} (\bm{q})=A_{{\rm \scriptscriptstyle P/M}} (-\bm{q})
\end{equation}
and similarly
\begin{equation}
   \label{eq:9}
   \tilde{t}^{*} (\bm{q})=\tilde{t} (-\bm{q}),
\end{equation}
because of the reality of the interaction potential. Translational
invariance of the interaction, leading to the invariance of $V_{{\rm
    \scriptscriptstyle PM}}$ under a global translation, is obvious
in~\eqref{eq:2} because the coupling through the potential only
depends on the relative positions of the two local operator densities,
and comes about in~\eqref{eq:6} because the operators in~\eqref{eq:7}
simply transform under a phase $\exp({\frac{i}{\hbar}\bm{q}\cdot\bm{a}})$
under a translation of step $\bm{a}$. The relationship
between~\eqref{eq:2} and~\eqref{eq:6} can be most easily seen by
analogy with the following identity exploiting the fact that the
Fourier transform is a unitary transformation
\begin{equation}
   \label{eq:10}
   \langle f| v*g \rangle=
   \langle \tilde{f}| \widetilde{v*g} \rangle=
   \langle \tilde{f}|\tilde{v}\tilde{g} \rangle=\int d^3 \! \bm{q} \,
   \tilde{v}(\bm{q})\tilde{f}(-\bm{q})\tilde{g}(\bm{q}) ,
\end{equation}
where $f$, $v$ and $g$ are real functions, $\tilde{f}$ denotes the
Fourier transform and $*$ the
convolution product.
\par
We will now consider two general types of physically meaningful
translationally invariant couplings, corresponding to quite distinct
situations. The first is a density-density coupling, given by the
identifications
\begin{equation}
   \label{eq:11}
   A_{{\rm \scriptscriptstyle P}} (\bm{x})=\varphi^{\scriptscriptstyle \dagger}
   (\bm{x})\varphi (\bm{x})\equiv N_{{\rm \scriptscriptstyle P}} (\bm{x}),
\end{equation}
$N_{{\rm \scriptscriptstyle P}} (\bm{x})$ being the number-density
operator for the test particles, and
\begin{equation}
   \label{eq:12}
   A_{{\rm \scriptscriptstyle M}} (\bm{x})=\psi^{\scriptscriptstyle \dagger}
   (\bm{x})\psi (\bm{x})\equiv N_{{\rm \scriptscriptstyle M}} (\bm{x})
\end{equation}
respectively. One therefore has
\begin{equation}
   \label{eq:13}
    V_{{\rm \scriptscriptstyle PM}}=
\int d^3 \! \bm{x} \int d^3 \! \bm{y} \,
N_{{\rm \scriptscriptstyle P}} (\bm{x})
t (\bm{x} - \bm{y})
N_{{\rm \scriptscriptstyle M}} (\bm{y}) ,  
\end{equation}
or equivalently setting
\begin{equation}
   \label{eq:14}
   A_{{\rm \scriptscriptstyle P}} (\bm{q})=\int d^3 \! \bm{x} \,
e^{-\frac{i}{\hbar}\bm{q}\cdot\bm{x}} N_{{\rm \scriptscriptstyle P}}
(\bm{x})=
\int \frac{d^3 \! \bm{k}}{(2\pi\hbar)^{3}} \, a_{\bm{k}}^{\scriptscriptstyle \dagger}a_{\bm{k}+\bm{q}}
\end{equation}
and introducing the $\bm{q}$-component of the number-density operator
$\rho_{\bm{q}}$~\cite{Lovesey,Stringari}
\begin{equation}
   \label{eq:15}
    A_{{\rm \scriptscriptstyle M}} (\bm{q})=\int d^3 \! \bm{x} \,
e^{-\frac{i}{\hbar}\bm{q}\cdot\bm{x}} N_{{\rm \scriptscriptstyle M}}
(\bm{x})=
\int \frac{d^3 \! \bm{\eta}}{(2\pi\hbar)^{3}} \,
b_{\bm{\eta}}^{\scriptscriptstyle \dagger}b_{\bm{\eta}+\bm{q}}\equiv    \rho_{\bm{q}}
\end{equation}
the alternative expression
\begin{equation}
   \label{eq:16}
   V_{{\rm \scriptscriptstyle PM}}=
\int d^3 \! \bm{q} \,
\tilde{t} (\bm{q})
A_{{\rm \scriptscriptstyle P}} (\bm{q})
\rho_{\bm{q}}^{\scriptscriptstyle \dagger} (\bm{q}).
\end{equation}
Note that an interaction of the form~\eqref{eq:13} or
equivalently~\eqref{eq:16}, besides being translationally invariant,
commutes with the number operators $N_{{\rm \scriptscriptstyle P}}$ and $N_{{\rm
    \scriptscriptstyle M}}$, so that the elementary interaction events
do bring in exchanges of momentum between the test particle and the
environment, but the number of particles or quanta in both systems are
independently conserved, thus typically describing an interaction in
terms of collisions. 
\par
The other type of interaction we shall consider
is a density-displacement coupling, corresponding to the expressions
\begin{equation}
   \label{eq:17}
    A_{{\rm \scriptscriptstyle P}} (\bm{x})=\varphi^{\scriptscriptstyle \dagger}
   (\bm{x})\varphi (\bm{x})\equiv N_{{\rm \scriptscriptstyle P}} (\bm{x}),  
\end{equation}
as \eqref{eq:11} above for the particle, and
\begin{equation}
   \label{eq:18}
       A_{{\rm \scriptscriptstyle M}} (\bm{x})=
\int \frac{d^3 \! \bm{\eta}}{(2\pi\hbar)^{3}} \,
(b_{\bm{\eta}}+b_{-\bm{\eta}}^{\scriptscriptstyle \dagger})
e^{\frac{i}{\hbar}\bm{\eta}\cdot\bm{x}}
\equiv   u(\bm{x})
\end{equation}
for the macroscopic system, where $u(\bm{x})$ is often called
displacement operator~\cite{Ashcroft,SchwablQMII}, thus leading to
\begin{equation}
   \label{eq:19}
   V_{{\rm \scriptscriptstyle PM}}=
\int d^3 \! \bm{x} \int d^3 \! \bm{y} \,
N_{{\rm \scriptscriptstyle P}} (\bm{x})
t (\bm{x} - \bm{y})
u(\bm{y}) ,   
\end{equation}
or in terms of the Fourier transformed quantities~\eqref{eq:14} and
\begin{equation}
   \label{eq:20}
   A_{{\rm \scriptscriptstyle M}} (\bm{q})=b_{\bm{q}}+b_{-\bm{q}}^{\scriptscriptstyle \dagger}=u(\bm{q}),
\end{equation}
to the equivalent expression
\begin{equation}
   \label{eq:21}
   V_{{\rm \scriptscriptstyle PM}}=
\int d^3 \! \bm{q} \,
\tilde{t} (\bm{q})
A_{{\rm \scriptscriptstyle P}} (\bm{q})
u^{\scriptscriptstyle \dagger} (\bm{q})=
\int d^3 \! \bm{q} \,
\tilde{t} (\bm{q})
A_{{\rm \scriptscriptstyle P}} (\bm{q})
(b_{\bm{q}}+b_{-\bm{q}}^{\scriptscriptstyle \dagger}).
\end{equation}
Contrary to~\eqref{eq:13} or~\eqref{eq:16}, the interaction considered
in~\eqref{eq:19} or~\eqref{eq:21} does not preserve the number of
quanta of the macroscopic system and rather than a collisional
interaction describes, e.g., a Fr\"ohlich-type interaction between
electron and phonon~\cite{Mahan}.
\par
Before showing the relationship between the above introduced
translationally invariant interactions and corresponding structures of
master-equation in the Markovian, weak-coupling limit, we briefly
discuss the connection with the most famous Caldeira Leggett model for
the quantum description of dissipation and decoherence. Despite, or
equivalently because of, its widespread use and relevance in
applications, it is well worth trying to elucidate the basic physics
behind the model, at least restricted to specific situations. In the
standard formulation of the Caldeira Leggett model (see for
example~\cite{Petruccione,IngoldLNPH02,Weiss99}) the Hamiltonian for
the environment is given in first quantization by the expression
\begin{equation}
   \label{eq:22}
   H_{{\rm \scriptscriptstyle M}}=\sum_{i=1}^{N}
\left( 
\frac{p_i^2}{2 m_i} + \frac{1}{2} m_i \omega_i^2 x_i^2
\right),
\end{equation}
which should describe a set of independent harmonic oscillators, while
the interaction term is given by (here and in the following we denote
one-particle operators referring to the test particle with a hat)
\begin{equation}
   \label{eq:23}
   V_{{\rm \scriptscriptstyle PM}}=-{\hat{\mathsf{x}}}\sum_{i=1}^{N}c_i x_i +
   {\hat{\mathsf{x}}}^2\sum_{i=1}^{N}\frac{c_i^2}{2 m_i\omega_i^2}, 
\end{equation}
typically focusing on a one-dimensional system, where the first term
is a position-position coupling and the second one is justified as a
counter-term necessary in order to restore the physical frequencies of
the dynamics of the microsystem, given e.g. by a Brownian particle. In
the absence of an external potential for the test particle it is also
observed that translational invariance, explicitly broken
by~\eqref{eq:22} and~\eqref{eq:23}, can be recovered by suitably
fixing the otherwise arbitrary coupling constants $c_i$ to be given
by~\cite{IngoldLNPH02}
\begin{equation}
   \label{eq:24}
   c_i=m_i\omega_i^2,
\end{equation}
which should not affect the relevant results which actually only
depend on the so-called spectral density
\begin{equation}
   \label{eq:25}
   J (\omega)=\sum_{i=1}^{N}\frac{c_i^2}{2 m_i\omega_i} \delta (\omega-\omega_i),
\end{equation}
which as a matter of fact is phenomenologically fixed. Since the
original idea behind the model is to give an effective description of
quantum dissipation in which the phenomenological quantities are to be
fixed by comparison with the classical model, thus working in a
semiclassical spirit, recovery of quantum Brownian motion in the sense
of Einstein in the case of a test particle in a homogeneous medium is
a natural requirement, and in fact the master-equation obtained from
the Caldeira Leggett model with the Ohmic prescription
for~\eqref{eq:25} is considered as the standard quantum description of
Brownian motion. Nonetheless, as stressed in~\cite{SolsPhysicaA94},
despite the aforementioned ad hoc adjustments the Caldeira Leggett
model does not comply with one of the basic features of Brownian
motion, i.e., translational invariance, and in fact also previous work
has focused on how to recover translational invariance in the quantum
description of dissipation~\cite{Gallis93}. In their analysis the
authors of~\cite{SolsPhysicaA94} try to recover a modified,
translationally invariant version of the Caldeira Leggett model by
exploiting a suitable limit of an interaction of the
density-displacement type considered above in~\eqref{eq:19} or
equivalently~\eqref{eq:21}. While this model might be the correct one
for other physical systems, we claim the Einstein's quantum Brownian
motion corresponds to a density-density coupling and we now see how the
Caldeira Leggett model is related to the long-wavelength limit of a
density-density coupling. Let us in fact consider~\eqref{eq:13}
and~\eqref{eq:16} restricting the expressions to the one-particle
sector for the test particle and to the $N$-particle sector for the
macroscopic system, thus obtaining, using a first quantization
formalism as in the Caldeira Leggett model,
\begin{equation}
   \label{eq:26}
   V_{{\rm \scriptscriptstyle PM}}=\sum_{i=1}^{N} t
   ({\hat{\mathsf{x}}}-\bm{x}_i)=
\int d^3 \! \bm{q} \,
\tilde{t} (\bm{q})
\sum_{i=1}^{N}  e^{-\frac{i}{\hbar}\bm{q}\cdot({\hat{\mathsf{x}}}-\bm{x}_i)}.
\end{equation}
Considering only small momentum transfers and thus taking the
long-wavelength limit of the expression, corresponding to a collective
response of the macroscopic medium, one obtains up to second order
\begin{equation}
   \label{eq:27}
   V_{{\rm \scriptscriptstyle PM}}
\, {\buildrel {\rm \scriptscriptstyle LWL}  \over {\approx}} \, 
N \int d^3 \! \bm{q} \, \tilde{t} (\bm{q})
%\\
- \frac{1}{2\hbar^2} \int d^3 \! \bm{q} \, \tilde{t} (\bm{q})
\sum_{i=1}^{N} [\bm{q}\cdot({\hat{\mathsf{x}}}-\bm{x}_i)]^2
+ O (q^4),
\end{equation}
where the term linear in $\bm{q}$ has dropped because of inversion
symmetry. Further exploiting isotropy, so that $\tilde{t}
(\bm{q})=\tilde{t} (q)$ and recalling the relationships
\begin{equation}
   \label{eq:28}
   \int d^3 \! \bm{q} \, \tilde{f} (\bm{q})=
\left.
f (\bm{x})
\right|_{\bm{x}=0}
\quad \mathrm{and} \quad
\int d^3 \! \bm{q} \,  q_i^2 \tilde{f} (\bm{q})=-\hbar^2
\left.
\frac{\partial^2 f}{\partial x_i^2}(\bm{x})
\right|_{\bm{x}=0},
\end{equation}
one has
\begin{equation}
   \label{eq:29}
   V_{{\rm \scriptscriptstyle PM}}
\, {\buildrel {\rm \scriptscriptstyle LWL}  \over {\approx}} \, 
N t (0)-\frac{1}{3}\Delta_2t (0)\, {\hat{\mathsf{x}}}\cdot\sum_{i=1}^{N}
\bm{x}_i 
%\\
+\frac{1}{6}\Delta_2t (0)\,\sum_{i=1}^{N}\bm{x}_i^2
+\frac{N}{6}\Delta_2t (0)\,{\hat{\mathsf{x}}}^2+ O (q^4).
\end{equation}
Here one easily recognizes the Caldeira Leggett model, though with
some constraints and modifications. First of all, as evident
from~\eqref{eq:26} and also stressed in~\cite{SolsPhysicaA94},
translational invariance is preserved in the long-wavelength limit
only provided that all terms up to a given order in $\bm{q}$ are
consistently kept, and this also applies to any calculation put
forward by means of~\eqref{eq:29}. This explains the appearance of the
so-called counter-term in~\eqref{eq:23}, as well as the
relationship~\eqref{eq:24} required in order to apparently restore
translational invariance. The symmetry requirement thus strictly fixes
the relationship between coefficients. However a position-position
coupling such as the one appearing in~\eqref{eq:29} is the common
feature of the long-wavelength limit of a density-density coupling
with a generic, not necessarily harmonic, potential. In the case in
which the potential is harmonic, i.e.
\begin{equation}
   \label{eq:30}
   t (\bm{x})=\frac{1}{2}m\omega^2 \bm{x}^2,
\end{equation}
one obtains from~\eqref{eq:29}
\begin{equation}
   \label{eq:31}
   V_{{\rm \scriptscriptstyle PM}}
\, {\buildrel {\rm \scriptscriptstyle LWL}  \over {\approx}} \, 
   \frac{1}{2}m\omega^2 \sum_{i=1}^{N} (\bm{x}_i^2 + {\hat{\mathsf{x}}}^2)-
m\omega^2 {\hat{\mathsf{x}}}\cdot\sum_{i=1}^{N} \bm{x}_i
\end{equation}
as in~\cite{IngoldLNPH02}. Let us note how in~\eqref{eq:29} the test
particle couples to the collective coordinate
\begin{equation}
   \label{eq:32}
   \bm{X}=\sum_{i=1}^{N} \bm{x}_i
\end{equation}
of the macroscopic system, proportional to its center of mass. In a
truly quantum picture of Einstein's Brownian motion, the gas has to be
described by identical particles (or mixtures thereof), so that one
cannot introduce different masses and different coupling
constants. According to~\eqref{eq:16} or~\eqref{eq:26} in a
density-density interaction the test particle is differently coupled
to the various $\bm{q}$-components of the number-density operator for
the macroscopic system $\rho_{\bm{q}}$, depending on the specific expression
of the interaction potential $t (\bm{x})$. Of course this is no more
relevant when interpreting the harmonic oscillators as representatives
of possible modes of the macroscopic system. Here and in the following
we are not aiming at a general critique of the Caldeira Leggett model,
which obviously has big merits, let alone its historical meaning as a
pioneering work in research on quantum dissipation. Rather, focusing
on the particular and at the same time paradigmatic example of the
quantum description of Einstein's Brownian motion, we want to put into
evidence the possible detailed microscopic physics behind the model,
especially in view of natural symmetry requirements, thus also opening
the way for alternative ways to look at and cope with dissipation and
decoherence in quantum mechanics, especially overcoming the limitation
to Gaussian statistics inherent in the Caldeira Leggett model. The
relevance that the microphysical coupling actually has in determining
which physical phenomena can be correctly described by a given model
has also been stressed in~\cite{AlickiOSID04}, where an analysis is
made of pure decoherence without dissipation, indicating that a full
density-density coupling rather than a position-position coupling as
in the Caldeira Leggett model (in the paper correctly formalized in
terms of a Bose field) should provide the proper way to describe pure,
recoilless decoherence.
\subsection{Structure of translation-covariant quantum-dynamical
  semigroups}
\label{sec:struct-transl-covar}
We now come back to the translationally invariant interactions given
by~\eqref{eq:13} and~\eqref{eq:16}, or~\eqref{eq:19}
and~\eqref{eq:21}, showing the master-equations they lead to in the
Markovian, weak-coupling limit. To do this we first observe that
because of homogeneity of the underlying medium and translational
invariance of the interaction potential, the reduced dynamics of the
test particle must also be invariant under translations, so that the
generator of the quantum-dynamical semigroup driving the dynamics of
the test particle, i.e., giving the master-equation, must comply with
the general characterization of translation-covariant generators of
quantum-dynamical semigroups given in recent, seminal work by
Holevo~\cite{HolevoRMP32,HolevoRMP33,HolevoRAN,HolevoJMP}. Given the
unitary representation ${\hat{\mathsf{U}}}
(\bm{a})=\exp({-\frac{i}{\hbar}\bm{a}\cdot{\hat{\mathsf{p}}}})$,
$\bm{a}\in \mathbb{R}^3$ of the group of translations $\mathbb{R}^3$
in the test particle Hilbert space, a mapping $\mathcal{L}$ acting on
the statistical operators in this space is said to be
translation-covariant if it commutes with the action of the unitary
representation, i.e.
\begin{equation}
   \label{eq:33}
   \mathcal{L}[{\hat{\mathsf{U}}}
(\bm{a}){\hat \varrho}{\hat{\mathsf{U}}}^{\scriptscriptstyle \dagger}
(\bm{a})]={\hat{\mathsf{U}}}
(\bm{a})\mathcal{L}[{\hat \varrho}]{\hat{\mathsf{U}}}^{\scriptscriptstyle \dagger}
(\bm{a}),
\end{equation}
for any statistical operator ${\hat \varrho}$ and any translation
$\bm{a}$. Needless to say the notion of covariance under a given
symmetry group has proved very powerful not only in characterizing
mappings such as quantum-dynamical semigroups and operations, but also
observables, especially in the generalized sense of
POVM~\cite{Grabowski,HolevoNEW}. In the specific case of generators of
translation-covariant quantum-dynamical semigroups the result of
Holevo, while obviously fitting in the general framework set by the
famous Lindblad result~\cite{GoriniJMP76,Lindblad}, goes beyond it
giving much more detailed information on the possible structure of
operators appearing in the Lindblad form, information conveyed by
the symmetry requirements and relying on a quantum generalization of
the Levy-Kintchine formula. Referring to the papers by Holevo for the
related mathematical details (see also~\cite{vienna} for a brief
r\'esum\'e), the physically relevant structure of the generator is
given by
\begin{equation}
   \label{eq:34}
   \mathcal{L}[{\hat \varrho}]=-\frac{i}{\hbar}[H
   ({\hat{\mathsf{p}}}),{\hat \varrho}]
+\mathcal{L}_{G}[{\hat \varrho}]+\mathcal{L}_{P}[{\hat \varrho}],
\end{equation}
with $H ({\hat{\mathsf{p}}})$ a self-adjoint operator which is only a
function of the momentum operator of the test particle; the so-called
Gaussian part $\mathcal{L}_{G}$ is given by
\begin{align}
   \label{eq:35}
   \mathcal{L}_{G}[{\hat \varrho}]
=&-{i \over \hbar}
        \left[{\hat{\mathsf{y}}}_0+
        H_{\mathrm{\scriptscriptstyle eff}} ({\hat{\mathsf{x}}},{\hat{\mathsf{p}}})
        ,{\hat \varrho}
        \right]
\\ \nonumber
&+\sum_{k=1}^{r}
\left[K_k{\hat \varrho}K_k^{\dagger} -\frac{1}{2}\left\{K_k^{\dagger}K_k,{\hat \varrho}\right\} \right],
\end{align}
where
\begin{align*}
K_k &={\hat{\mathsf{y}}}_k+L_k ({\hat{\mathsf{p}}}) ,
\\
 {\hat{\mathsf{y}}}_k &=\sum_{i=1}^{3}a_{ki}{\hat{\mathsf{x}}}_i \quad
  k=0,\ldots, r\leq 3 \quad a_{ki}\in \mathbb{R},
\\
H_{\mathrm{\scriptscriptstyle eff}}
({\hat{\mathsf{x}}},{\hat{\mathsf{p}}})&=\frac{\hbar}{2i}\sum_{k=1}^{r}
({\hat{\mathsf{y}}}_k L_k ({\hat{\mathsf{p}}}) -L_k^{\dagger} ({\hat{\mathsf{p}}}){\hat{\mathsf{y}}}_k)
\end{align*}
and the remaining Poisson part takes the form
\begin{equation}
   \label{eq:36}
   \mathcal{L}_{P}[{\hat \varrho}]=
\int d\mu (\bm{q})\sum_{j=1}^{\infty}
\left[e^{{i\over\hbar}\bm{q}\cdot{\hat{\mathsf{x}}}}
L_j(\bm{q},{\hat{\mathsf{p}}})
{\hat \varrho}
L^{\dagger}_j(\bm{q},{\hat{\mathsf{p}}})e^{-{i\over\hbar}\bm{q}\cdot{\hat{\mathsf{x}}}}
%\right.
%\\
%\left.
-        \frac 12
        \left \{
        L^{\dagger}_j(\bm{q},{\hat{\mathsf{p}}})L_j(\bm{q},{\hat{\mathsf{p}}}),{\hat \varrho} 
        \right \}
\right],
\end{equation}
with $d\mu (\bm{q})$ a positive measure, ${\hat{\mathsf{x}}}$ and
${\hat{\mathsf{p}}}$ position and momentum operators for the test
particle respectively.  As it can be seen the
characterization is quite powerful, so that the only freedom left is
in the choice of a few coefficients and functions of the momentum
operator of the test particle ${\hat{\mathsf{p}}}$. These can be fixed
either referring to microphysical calculations, or relying on a suitably
guessed phenomenological Ansatz. In this kind of reduced dynamics the
information on the macroscopic system the test particle is interacting
with is essentially encoded in a suitable, possibly operator-valued,
two-point correlation function of the macroscopic system appearing
in the formal Lindblad structure. The key physical point is then the
identification of the relevant two-point correlation function,
depending both on the coupling between test particle and reservoir,
and on a characterization of the equilibrium state of the reservoir.
\subsection{Physical examples}
\label{sec:physical-examples}
The case of density-density coupling given by~\eqref{eq:13}
and~\eqref{eq:16}, when the reservoir is given by a free quantum gas,
has been dealt with in~\cite{art3,art4,art5}, and the relevant test
particle correlation function turns out to be the so-called dynamic
structure factor~\cite{Lovesey,Stringari}
\begin{equation}
   \label{eq:37}
   S (\bm{q},E)=\frac{1}{2\pi\hbar}\frac{1}{N}        
\int dt \,
        e^ {{i\over\hbar}Et}
         \langle  \rho_{\bm{q}}^{\scriptscriptstyle \dagger}\rho_{\bm{q}} (t) \rangle,
\end{equation}
which can be written in an equivalent way as
\begin{equation}
   \label{eq:38}
    S (\bm{q},E)=\frac{1}{N}\sum_{mn}\frac{e^{-\beta
        E_n}}{\mathcal{Z}}
|\langle m |\rho_{\bm{q}}| n\rangle|^2 \delta (E+E_m-E_n),
\end{equation}
where contrary to the usual conventions, momentum and energy
are considered to be positive when transferred to the test particle,
on which we are now focusing our attention, rather than on the
macroscopic system. The master-equation then takes the form
%\begin{widetext}
\begin{align}
   \label{eq:39}
   \frac{d{\hat \varrho}}{dt}=
        &{}-
        {i \over \hbar}[
        {\hat{\mathsf{H}}}_0
        ,
        {\hat \varrho}
        ]
\\ \nonumber
        &{}+
        {2\pi \over\hbar}
        (2\pi\hbar)^3
        n
        \int d^3\!
        \bm{q}
        \,  
        {
        | \tilde{t} (q) |^2
        }
        \Biggl[
        e^{{i\over\hbar}\bm{q}\cdot{\hat{\mathsf{x}}}}
        \sqrt{
        S(\bm{q},E (\bm{q},{\hat{\mathsf{p}}}))
        }
        {\hat \varrho}
        \sqrt{
        S(\bm{q},E (\bm{q},{\hat{\mathsf{p}}}))
        }
        e^{-{i\over\hbar}\bm{q}\cdot{\hat{\mathsf{x}}}}
        -
        \frac 12
        \left \{
        S(\bm{q},E (\bm{q},{\hat{\mathsf{p}}})),
        {\hat \varrho}
        \right \}
        \Biggr],
\end{align}
%\end{widetext}
where ${\hat{\mathsf{H}}}_0$ is the free particle Hamiltonian, $n$ the
density of the homogeneous gas, and the dynamic structure factor
appears operator-valued: in fact the energy transfer in each
collision, which is given by
\begin{equation}
   \label{eq:40}
   E (\bm{q},\bm{p})=\frac{(\bm{p}+\bm{q})^2}{2M}-\frac{\bm{p}^2}{2M},
\end{equation}
with $M$ the mass of the test particle, is turned into an operator by
replacing $\bm{p}$ with ${\hat{\mathsf{p}}}$.
For the case of a free gas of particles obeying Maxwell-Boltzmann
statistics the dynamic structure factor takes the explicit form
\begin{equation}
   \label{eq:41}
         S_{\rm \scriptscriptstyle MB}(\bm{q},E)
        =
        \sqrt{\frac{\beta m}{2\pi}}        
        {
        1
        \over
        q
        }
       e^{-{
        \beta
        \over
             8m
        }
        {
        (2mE + q^2)^2
        \over
                  q^2
        }}
\end{equation}
with $\beta$ the inverse temperature and $m$ the mass of the gas
particles. A density-displacement type of coupling as in~\eqref{eq:19}
and~\eqref{eq:21} has been dealt with in~\cite{SpohnJSP77,SpohnRMP},
considering an environment essentially given by a phonon bath. The
relevant test particle correlation function in these kind of models is
given by the following spectral function~\cite{Lovesey,SchwablQMII}
\begin{equation}
   \label{eq:42}
   S (\bm{q},E)=\frac{1}{2\pi\hbar}
\int dt \,
        e^ {{i\over\hbar}Et}
         \langle  u^{\scriptscriptstyle \dagger}(\bm{q})u(\bm{q},t) \rangle,   
\end{equation}
given by a linear combination of correlation functions of the form
\begin{equation}
   \label{eq:43}
   A(\bm{q},E)=\frac{1}{2\pi\hbar}
\int dt \,
        e^ {{i\over\hbar}Et}
         \langle  b_{\bm{q}}^{\scriptscriptstyle \dagger}b_{\bm{q}} (t) \rangle,
\end{equation}
which can also be written~\cite{Griffin}
\begin{equation}
   \label{eq:44}
     A(\bm{q},E)=\sum_{mn}\frac{e^{-\beta
        E_n}}{\mathcal{Z}}
|\langle m |b_{\bm{q}}| n\rangle|^2 \delta (E+E_m-E_n).
\end{equation}
Contrary to the smooth expression
of the dynamic structure factor for a free quantum gas given
in~\eqref{eq:41}, the spectral function~\eqref{eq:43} has the highly
singular structure
\begin{equation}
   \label{eq:45}
   S (\bm{q},E)=[1+N_{\beta} (\hbar\omega_{\bm{q}})]\delta
   (E+\hbar\omega_{\bm{q}})+N_{\beta} (\hbar\omega_{\bm{q}})\delta
   (E-\hbar\omega_{\bm{q}}), 
\end{equation}
with
\begin{equation}
   \label{eq:46}
   N_{\beta} (\hbar\omega_{\bm{q}})=\frac{1}{e^{\beta\hbar\omega_{\bm{q}}}-1},
\end{equation}
where the exact frequencies $\hbar\omega_{\bm{q}}$ of the phonon
appear. The smooth energy dependence of the test particle correlation
function used in the derivation of~\eqref{eq:39}, allowing an exact
treatment in the case of a free gas of Maxwell-Boltzmann particles,
here no longer applies, and in fact the master-equation has only been
worked out for the diagonal matrix elements of the statistical
operator in the momentum representation. Setting $\varrho
(\bm{p})\equiv \langle \bm{p}|{\hat \varrho}|\bm{p}\rangle$ one has
\begin{equation}
   \label{eq:47}
   \frac{d\varrho}{dt}(\bm{p})=\int d^3 \! \bm{q} \,
| \tilde{t} (q) |^2
        \Biggl[
        S(\bm{q},E (\bm{q},\bm{p}-\bm{q}))
        \varrho (\bm{p}-\bm{q})
        -
        S(\bm{q},E (\bm{q},\bm{p}))\varrho (\bm{p})
        \Biggr],
\end{equation}
using the notation introduced in~\eqref{eq:40}. One immediately sees
that both~\eqref{eq:39} and~\eqref{eq:47} fit in the general
expression~\eqref{eq:36} for the Poisson part of the generator of a
translation-covariant quantum-dynamical semigroup given by Holevo,
with the $|L_j(\bm{q},{\hat{\mathsf{p}}})|^2$ operators replaced by the
spectral functions~\eqref{eq:37} and~\eqref{eq:42} respectively, the
integration measure $d\mu (\bm{q})$ corresponding to the
Lebesgue measure with a weight given by the square modulus of
the Fourier transform of the interaction potential. 
It is here already apparent that the presented results~\eqref{eq:39}
and~\eqref{eq:47}, pertaining to the Poisson part~\eqref{eq:36} of
the general structure of generator of a translation-covariant
quantum-dynamical semigroup~\eqref{eq:34}, go beyond the limitation to
Gaussian statistics typical of the Caldeira Leggett model.
\par
The relevant correlation function for these translation-covariant
master-equations thus appears to be given by the Fourier transform with respect
to energy of the time-dependent autocorrelation function of the
operator of the macroscopic system appearing in the interaction
potential $V_{{\rm \scriptscriptstyle PM}}$ when written in the
form~\eqref{eq:6}, i.e.
\begin{equation}
   \label{eq:48}
      S (\bm{q},E)=\frac{1}{2\pi\hbar}
\int dt \,
        e^ {{i\over\hbar}Et}
         \langle  A_{{\rm \scriptscriptstyle M}}^{\scriptscriptstyle
           \dagger}(\bm{q}) A_{{\rm \scriptscriptstyle M}}(\bm{q},t)
         \rangle.    
\end{equation}
The parameter $\bm{q}$ one integrates over in~\eqref{eq:36}, with a
weight given by the square modulus of the Fourier transform
of the interaction potential appearing in~\eqref{eq:6}, is to be seen
as an element of the translation group, physically corresponding to
the possible momentum transfers in the single collisions. The key
difference between the two models lies in the physical meaning of the
different correlation functions. The dynamic structure
factor~\eqref{eq:37} is linked to the so-called density fluctuations
spectrum, accounting for particle number conservation of the
macroscopic system. This connection to density fluctuations brings
into play the other key feature of Einstein's Brownian motion,
i.e., the molecular, discrete nature of matter. As we shall see shortly,
the smooth correlation function arising in connection with this
density-density coupling allows us to take a diffusive limit of the
reduced dynamics, thus obtaining the quantum description of Einstein's
Brownian motion. On the contrary in~\eqref{eq:42} the typically
quantized spectrum of a harmonic oscillator appears, thus leading to
the singular function~\eqref{eq:45}, so that as stressed
in~\cite{SpohnRMP} rather than a diffusion equation one necessarily has
a jump process.
\section{Fluctuation-dissipation theorem}
\label{sec:fluct-diss-theor}
In the previous paragraph we have tried to point out and analyze the
typical structures for the quantum description of dissipation and
decoherence in the Markovian case that come into play when the first
of the two key features of Einstein's Brownian motion mentioned in
Sect.~\ref{sec:introduction} is taken into account, i.e., translational
invariance. We now focus on the second key feature, i.e., the
connection with the discrete nature of matter, which Einstein actually
wanted to demonstrate. As already hinted at the end of
Sect.~\ref{sec:transl-invar}, in the present paper we substantiate the
claim that the correct description of Einstein's Brownian motion is
obtained considering a density-density coupling. As we shall see in
detail in Sect.~\ref{sec:quant-descr-einst} this happens thanks to the
fact that the two-point correlation function appearing in the
master-equation in this case is the dynamic structure
factor~\eqref{eq:37}, where the Fourier transform of the
number-density operator $\rho_{\bm{q}}$, as given in~\eqref{eq:15},
appears. This function is in fact directly related to the density
fluctuations in the medium, as it can be seen writing it, rather than
in the form~\eqref{eq:37}, relevant for the comparison between the
different types of translational invariance interactions and related
master-equations, in the following way~\cite{Lovesey}:
\begin{equation}
   \label{eq:49}
   S (\bm{q},E)=\frac{1}{2\pi\hbar}
\int dt \int d^3 \! \bm{x} \, 
        e^ {\frac{i}{\hbar}(E t -
        \bm{q}\cdot\bm{x})}
G (\bm{x},t),
\end{equation}
i.e., as Fourier transform with respect to energy and momentum transfer
of the time dependent density correlation function 
\begin{equation}
   \label{eq:50}
   G (\bm{x},t)=\frac{1}{N}\int d^3 \! \bm{y} \,
        \left \langle  
         N_{{\rm \scriptscriptstyle M}}(\bm{y})  
         N_{{\rm \scriptscriptstyle M}}(\bm{x}+\bm{y},t)
         \right \rangle.
\end{equation}
Here the connection with density fluctuations and therefore discrete
nature of matter is manifest. Introducing the real correlation
functions
\begin{equation}
   \label{eq:51}
   \begin{split}
      \phi^{-} (\bm{q},t)&=\frac{i}{\hbar
        N}\langle[\rho_{\bm{q}}(t),\rho_{\bm{q}}^{\scriptscriptstyle
        \dagger}]\rangle
      \\
      \phi^{+} (\bm{q},t)&=\frac{1}{\hbar
        N}\langle\{\rho_{\bm{q}}(t),\rho_{\bm{q}}^{\scriptscriptstyle
        \dagger}\}\rangle,
\end{split}
\end{equation}
where $\{,\}$ denotes the anticommutator, the fluctuation-dissipation
theorem can be formulated in terms of the dynamic structure factor as
follows
\begin{equation}
   \label{eq:52}
   \begin{split}
      \phi^{-} (\bm{q},t)&=-\frac{2}{\hbar}\int^{0}_{-\infty} dE\,
      \sin \left( \frac{E}{\hbar}t\right)\left( 1-e^{\beta E}\right)S
      (\bm{q},E)
      \\
      \phi^{+} (\bm{q},t)&=-\frac{2}{\hbar}\int^{0}_{-\infty} dE\,
      \cos \left( \frac{E}{\hbar}t\right)\coth\left(
        \frac{\beta}{2}E\right) \left( 1-e^{\beta E}\right)S
      (\bm{q},E).
\end{split}
\end{equation}
We stress once again that contrary to the usual perspective in linear
response theory, we are here concerned with the reduced dynamics of
the test particle, so that we take as positive momentum and energy
transferred to the particle. The dynamic structure factor can also be
directly related to the dynamic response function
$\chi'' (\bm{q},E)$~\cite{Stringari}, according to
\begin{equation}
   \label{eq:53}
   \begin{split}
      S (\bm{q},E)&=\frac{1}{2\pi}\left[1- \coth\left(
          \frac{\beta}{2}E\right)\right]\chi'' (\bm{q},E)
      \\
      &=\frac{1}{\pi}\frac{1}{1-e^{\beta E}}\chi'' (\bm{q},E),
\end{split}
\end{equation}
the relationship leading to the important fact that while the dynamic
response function is an odd function of energy, the dynamic structure
factor obeys the so-called detailed balance condition 
\begin{equation}
   \label{eq:54}
   S (\bm{q},E)=e^{-\beta E}S (-\bm{q},-E),
\end{equation}
a property granting the existence of a stationary state for the
master-equation~\eqref{eq:39}, as shown in~\cite{art5}. In terms of the dynamic
response function the fluctuation-dissipation theorem can also be
written
\begin{equation}
   \label{eq:55}
   \begin{split}
      \phi^{-} (\bm{q},t)&=-\frac{2}{\pi\hbar}\int^{0}_{-\infty} dE\,
      \sin \left( \frac{E}{\hbar}t\right)\chi'' (\bm{q},E)
      \\
      \phi^{+} (\bm{q},t)&=-\frac{2\pi}{\hbar}\int^{0}_{-\infty} dE\,
      \cos \left( \frac{E}{\hbar}t\right)\coth\left(
        \frac{\beta}{2}E\right)\chi'' (\bm{q},E),
\end{split}
\end{equation}
a formulation that will prove useful for later comparison with the
Caldeira Leggett model. The most significant formulation of the so-called
fluctuation-dissipation theorem for the physics we are considering
is however neither~\eqref{eq:52} nor~\eqref{eq:55}, but is to be
traced back to a seminal paper by van
Hove~\cite{vanHove,SchwablQMII}. In fact he showed that the scattering
cross-section of a microscopic probe off a macroscopic sample can be
written in Born approximation in the following way
\begin{equation}
   \label{eq:56}
     \frac{d^2 \sigma}{d\Omega_{p'} dE_{p'}} (\bm{p}) =
\left({2\pi\hbar}\right)^6
\left(\frac{M}{2\pi\hbar^2}\right)^2
\frac{p'}{p}
        {
        | \tilde{t} (q) |^2
        }
  S (\bm{q},E)
,
\end{equation}
where a particle of mass $M$ changes its momentum from $\bm{p}$ to
$\bm{p}' = \bm{p}+\bm{q}$ scattering off a medium with
dynamic structure factor $S (\bm{q},E)$. This is the most pregnant 
formulation of the fluctuation-dissipation relationship for the case of a
test particle interacting through collisions with a macroscopic
fluid. The energy and momentum transfer to the particle, characterized
by the expression of the scattering cross-section at
l.h.s. of~\eqref{eq:56} are related to the density fluctuations of the
macroscopic fluid appearing through the dynamic structure factor at
r.h.s. of~\eqref{eq:56}. One of the basic ideas of Einstein's Brownian
motion, i.e., the discrete nature of matter, once again appears in the
formulation~\eqref{eq:56} of the fluctuation-dissipation relationship. From
the comparison between~\eqref{eq:56} and~\eqref{eq:39} one sees that
the reduced dynamics is actually driven by the collisional scattering
cross-section, in particular the last term of~\eqref{eq:39} can also
be written
\begin{equation}
   \label{eq:57}
   -\frac{n}{2M}\{|{\hat{\mathsf{p}}}|\sigma ({\hat{\mathsf{p}}}),{\hat \varrho}\},
\end{equation}
where $\sigma (\bm{p})$ is the total macroscopic scattering
cross-section obtained from the differential expression~\eqref{eq:56}
for a test particle with incoming momentum $\bm{p}$. The
term~\eqref{eq:57} can be seen quite naturally as a loss term in a
kinetic equation, and in fact~\eqref{eq:39} is actually to be seen as
a quantum version of the linear Boltzmann equation~\cite{art7}.
Besides this, from the direct relation~\eqref{eq:56} between
scattering cross-section and dynamic structure factor one reads on
physical grounds the positivity of the correlation function, a
property exploited in~\eqref{eq:39} in order to take the square root.
\par
We now compare the above formulations of the fluctuation-dissipation
theorem with the ones encountered in the long-wavelength limit of the
density-density coupling type of translationally invariant
interaction, which as shown in Sect.~\ref{sec:transl-invar} is strongly
related to the Caldeira Leggett model. In the long-wavelength limit
the $\bm{q}$-component of the number-density operator becomes
\begin{equation}
   \label{eq:58}
\rho_{\bm{q}}
\, {\buildrel {\rm \scriptscriptstyle LWL}  \over {\approx}} \,  
N -\frac{i}{\hbar} \bm{q}\cdot\sum_{i=1}^{N}
\bm{x}_i+ O (q^2),
\end{equation}
and once again the collective coordinate $\bm{X}=\sum_{i=1}^{N}
\bm{x}_i$ introduced in~\eqref{eq:32} is put into evidence. The relevant correlation
functions then become
\begin{equation}
   \label{eq:59}
   \begin{split}
      \phi^{-}_{ij} (\bm{q},t)&=\frac{i}{\hbar
        N}\langle[\bm{X}_{i}(t),\bm{X}_{j}]\rangle
      \\
      \phi^{+}_{ij} (\bm{q},t)&=\frac{1}{\hbar
        N}\langle\{\bm{X}_{i}(t),\bm{X}_{j}\}\rangle,
\end{split}
\end{equation}
the indexes $i$ and $j$ here denoting Cartesian components of the
collective coordinate~\eqref{eq:32}. Introducing accordingly the
spectral function
\begin{equation}
   \label{eq:60}
   S_{ij} (E)=\frac{1}{2\pi\hbar}\frac{1}{N}        
\int dt \,
        e^ {{i\over\hbar}Et} \langle\bm{X}_{j} \bm{X}_{i}(t)\rangle,
\end{equation}
the fluctuation-dissipation theorem reads
\begin{equation}
   \label{eq:61}
   \begin{split}
      \phi^{-}_{ij} (t)&=-\frac{2}{\hbar}\int^{0}_{-\infty} dE\, \sin
      \left( \frac{E}{\hbar}t\right)\left( 1-e^{\beta E}\right) S_{ij}
      (E)
      \\
      \phi^{+}_{ij} (t) &=-\frac{2}{\hbar}\int^{0}_{-\infty} dE\, \cos
      \left( \frac{E}{\hbar}t\right)\coth\left(
        \frac{\beta}{2}E\right) \left( 1-e^{\beta E}\right) S_{ij}
      (E).
\end{split}
\end{equation}
With the help of the response function $\chi_{ij}'' (E)$
\begin{equation}
   \label{eq:62}
   S_{ij} (E)=\frac{1}{\pi}\frac{1}{1-e^{\beta E}}\chi_{ij}'' (E),
\end{equation}
the relations~\eqref{eq:61} can be also written as
\begin{equation}
   \label{eq:63}
   \begin{split}
      \phi^{-}_{ij} (t)&=-\frac{2}{\pi\hbar}\int^{0}_{-\infty} dE\,
      \sin \left( \frac{E}{\hbar}t\right)\chi_{ij}'' (E)
      \\
      \phi^{+}_{ij} (t) &=-\frac{2}{\pi\hbar}\int^{0}_{-\infty} dE\,
      \cos \left( \frac{E}{\hbar}t\right)\coth\left(
        \frac{\beta}{2}E\right) \chi_{ij}'' (E).
\end{split}
\end{equation}
While a formulation of the fluctuation-dissipation theorem like the van
Hove relation~\eqref{eq:56} is missing in this long-wavelength limit,
the relations \eqref{eq:63}, involving expectation values of
commutator and anticommutator of the components of the collective
coordinates, are the ones to be compared with the typical relations
used in order to introduce the so-called spectral
density~\eqref{eq:25} in the Caldeira Leggett model. In fact if all
coupling constants $c_i$ are put equal to $c$, as should be forced
upon in the case of Einstein's quantum Brownian motion, in which the
particle interacts through collisions with a collection of identical,
indistinguishable particles, the spectral density, when expressed in
terms of energy $E$ rather than frequency $\omega$, would be related
to the response function $\chi'' (E)$ for a one dimensional system
according to
\begin{equation}
   \label{eq:64}
   J (E)=\frac{c^2}{\pi}\chi'' (E).
\end{equation}
The relation~\eqref{eq:64}, first intuitively guessed in~\cite{CLAP83},
actually shows how in the friction coefficient, usually
phenomenologically introduced through the spectral density, features
of both the single interaction events and the reservoir do appear. In
Sect.~\ref{sec:quant-descr-einst} we will give a microscopic expression
for the friction coefficient in the case of Einstein's quantum
Brownian motion, in which both features do appear: the coupling
through the Fourier components of the interaction potential, and the
reservoir through certain values of the dynamic structure factor.
\section{Quantum description of Einstein's Brownian motion}
\label{sec:quant-descr-einst}
Relying on the premises of Sect.~\ref{sec:transl-invar}
and~\ref{sec:fluct-diss-theor} we now come to the master-equation for
the quantum description of Einstein's Brownian motion. The
requirement of translational invariance has been settled in
Sect.~\ref{sec:transl-invar}, while the connection between reduced
dynamics of the test particle and density fluctuations in the medium,
coming about because of its discrete nature, has been taken into
account in Sect.~\ref{sec:fluct-diss-theor}, considering a
density-density coupling and thus coming to~\eqref{eq:39}. The last
step to be taken is to consider the test particle much more massive
than the particles making up the gas, i.e., the Brownian limit $m/M \ll
1$, which in turn implies considering both small energy and momentum
transfers, similarly to the classical case~\cite{Uhlenbeck48}. We
therefore start from~\eqref{eq:39} and consider a free gas of
Maxwell-Boltzmann particles, so that taking the limiting expression
of~\eqref{eq:41} when the ratio between the masses is much smaller
than one, or equivalently considering small energy transfers, i.e.
\begin{equation}
   \label{eq:65}
           S^{\scriptscriptstyle\infty}_{\rm \scriptscriptstyle
          MB}(\bm{q},E) 
        =
        \sqrt{\frac{\beta m}{2\pi}}
        {
        1
        \over
        q
        }
        e^{
        -{
        \beta
        \over
             8m
        }
        q^2
        }
        e^{
        -\frac{\beta}{2}
        E
        },
\end{equation}
one obtains the master-equation~\cite{art3,art4,art5}
\begin{widetext}
\begin{align}
   \label{eq:66}
   \frac{d{\hat \varrho}}{dt}=
   &{}-
        {i \over \hbar}[
        {\hat{\mathsf{H}}}_0
        ,
        {\hat \varrho}
        ]
\\ \nonumber
        &{}+
        {2\pi \over\hbar}
        (2\pi\hbar)^3
        n
        \sqrt{\frac{\beta m}{2\pi}}
        \int d^3\!
        \bm{q}
        \,  
        \frac{| \tilde{t} (q) |^2}{q}
        e^{-\frac{\beta}{8m}\left( 1+2\frac{m}{M}\right) q^2}
      \Biggl[
        e^{{i\over\hbar}\bm{q}\cdot{\hat{\mathsf{x}}}}
        e^{-{\beta\over 4M}\bm{q}\cdot{\hat{\mathsf{p}}}}
        {\hat \varrho}
        e^{-{\beta\over 4M}\bm{q}\cdot{\hat{\mathsf{p}}}}
        e^{-{i\over\hbar}\bm{q}\cdot{\hat{\mathsf{x}}}}
        - {1\over 2}
        \left \{
        e^{-{\beta\over 2M}\bm{q}\cdot{\hat{\mathsf{p}}}}
        ,
        {\hat \varrho}
        \right \}
        \Biggr],
\end{align}
%\end{widetext}
which in the limit of small momentum transfer leads, of necessity as
can be seen from the Gaussian contribution in Holevo's
result~\eqref{eq:35} but also from previous
work~\cite{LindbladQBM,Sandulescu}, to a Caldeira Leggett type
master-equation, however without shortcomings related to the lack of
preservation of positivity of the statistical operator. The
master-equation takes the form
%\begin{widetext}
\begin{equation}
   \label{eq:67}
           {  
        d {\hat \varrho}  
        \over  
                dt  
        }  
        =
        -
        {i\over\hbar}
        [
        {{\hat{\mathsf{H}}}_0}
        ,{\hat \varrho}
        ]
        -
        {i\over\hbar}
        \frac{\eta}{2}
        \sum_{i=1}^3
        \left[  
        {\hat{\mathsf{x}}}_i ,
        \left \{  
        {\hat{\mathsf{p}}}_i,{\hat \varrho}
        \right \}  
        \right]         -
        {
        D_{pp}  
        \over
         \hbar^2
        }
        \sum_{i=1}^3
        \left[  
        {\hat{\mathsf{x}}}_i,
        \left[  
        {\hat{\mathsf{x}}}_i,{\hat \varrho}
        \right]  
        \right]  
        -
        {
        D_{xx}
        \over
         \hbar^2
        }
        \sum_{i=1}^3
        \left[  
        {\hat{\mathsf{p}}}_i,
        \left[  
        {\hat{\mathsf{p}}}_i,{\hat \varrho}
        \right]  
        \right]  
 ,
\end{equation}
\end{widetext}
with
\begin{equation}
   \label{eq:68}
   D_{pp}=\frac{M}{\beta}\eta\quad \text{and}\quad
   D_{xx}=\frac{\beta\hbar^2}{16 M}\eta .
\end{equation}
The friction coefficient $\eta$ is uniquely determined on the basis of
the microscopic information on interaction potential and correlation
function of the macroscopic system, according to
\begin{equation}
   \label{eq:69}
   \eta=        
\frac{\beta}{2M}
{2\pi \over\hbar}
        (2\pi\hbar)^3
        n
        \int d^3\!
        \bm{q}
        \,  
        | \tilde{t} (q) |^2 \,
        \frac{q^2}{3} S (\bm{q},E=0),
\end{equation}
the factor 3 being related to the space dimensions, or equivalently
\begin{equation}
   \label{eq:70}
      \eta=        
\frac{\beta}{2M}
{2\pi \over\hbar}
        (2\pi\hbar)^2
        n
        \int d^3\!
        \bm{q}
        \,  
        | \tilde{t} (q) |^2 \,
        \frac{q^2}{3}
\frac{1}{N}        
\int dt \,
\langle  \rho_{\bm{q}}^{\scriptscriptstyle \dagger}\rho_{\bm{q}} (t) \rangle,
\end{equation}
thus proving in a specific physical case of interest the so-called
standard wisdom expecting the decoherence and dissipation rate to be
connected with the value at zero energy of some suitable spectral
function~\cite{AlickiOSID04}. Introducing the Fourier transform of the
gradient of the number-density operator, which we indicate by $\nabla \rho_{\bm{q}}$
\begin{equation}
   \label{eq:71}
   \nabla \rho_{\bm{q}}\equiv \bm{q}\rho_{\bm{q}} = -i\hbar 
\int d^3 \! \bm{x}\,
e^{-\frac{i}{\hbar}\bm{q}\cdot\bm{x}} \nabla N_{{\rm \scriptscriptstyle M}}(\bm{x}),
\end{equation}
the friction coefficient can also be written in terms of the time
dependent autocorrelation function of $\nabla \rho_{\bm{q}}$ according to
\begin{equation}
   \label{eq:72}
       \eta=        
\frac{\beta}{6M}
{2\pi \over\hbar}
        (2\pi\hbar)^2
        n
        \int d^3\!
        \bm{q}
        \,  
        | \tilde{t} (q) |^2 \,
\frac{1}{N}        
\int dt \,
\langle  \nabla\rho_{\bm{q}}^{\scriptscriptstyle \dagger}\cdot\nabla\rho_{\bm{q}}
(t) \rangle.
\end{equation}
It is worth noticing how, contrary to the usual Caldeira Leggett
model, the friction coefficient will generally exhibit an explicit
temperature dependence, being related both to the expectation value of the
operators $\rho_{\bm{q}}$ and to the interaction potential. No energy cutoff
needs to be introduced, since all quantities appearing in the
calculations remain finite, being directly linked to the relevant
physical properties of the macroscopic system the test particle is
interacting with. Note that introducing the thermal momentum spread
\begin{equation}
   \label{a}
   {\Delta p}^2_{\rm
  \scriptscriptstyle th}=\frac{M}{\beta}
\end{equation}
and the square thermal wavelength
\begin{equation}
   \label{b}
   {\Delta x}^2_{\rm
  \scriptscriptstyle th}=\frac{\beta\hbar^2}{4M}
\end{equation}
satisfying the
minimum uncertainty relation
\begin{equation}
   \label{c}
   {\Delta p}_{\rm \scriptscriptstyle
  th}{\Delta x}_{\rm \scriptscriptstyle th}=\frac{\hbar}{2}
\end{equation}
the coefficients given in~\eqref{eq:68} can also be expressed in the
form
\begin{equation}
   \label{d}
   D_{pp}=\eta {\Delta p}^2_{\rm
  \scriptscriptstyle th}\quad \text{and}\quad
   D_{xx}=\frac{\eta}{4} {\Delta x}^2_{\rm
  \scriptscriptstyle th}.   
\end{equation}
\par
The main difference between~\eqref{eq:67} and the
master-equation introduced by Caldeira and Leggett for the description of
quantum Brownian motion, apart from the microphysical expression for
the appearing coefficients, lies in the appearance of the last
contribution, given by a double commutator with the momentum operator
of the Brownian particle, and corresponding to position
diffusion. This term, which here appears in the expansion for small
energy and momentum transfer of the dynamic structure factor, is
directly linked to preservation of positivity of the statistical
operator, and in fact in the past many different amendments of the
Caldeira Leggett master-equation have been proposed in the literature
introducing a term of this kind~\cite{art3,debate,reply}, even though
it is not obvious how to actually experimentally check the relevance
of this term, essentially quantum in origin, as can also be seen
from~\eqref{c} and~\eqref{d}. In recent work~\cite{art7} it has been
shown how this contribution might lead in the strong friction limit to
a typically quantum correction to Einstein's diffusion coefficient,
only relevant at low temperatures, thus opening the way to the
conception of future experiments in which to possibly check the
correction, as considered in~\cite{art11}.
\section{Conclusions and outlook}
\label{sec:conclusions-outlook}
In the present paper a fully quantum approach to the description of
Brownian motion in the sense of Einstein, i.e., considering a massive
test particle interacting through collisions with a background of much
lighter ones, has been presented. The two cardinal requirements
determining the quantum description of the reduced dynamics are
translational invariance and the connection with the discrete, atomistic nature
of the medium, along the lines of Einstein's original confrontation
with the problem. The former implies the choice of a translationally
invariant interaction potential and leads to the requirement of
translation covariance for the quantum-dynamical semigroup giving the
time evolution, according to Holevo's results~\cite{HolevoJMP} as seen
in Sect.~\ref{sec:transl-invar}; the latter relates the dynamics to
the density fluctuations in the fluid, expressed in terms of the
dynamic structure factor, first introduced by van Hove~\cite{vanHove},
and ensuring the physically most telling formulation of the
fluctuation-dissipation theorem for the considered case, as seen in
Sect.~\ref{sec:fluct-diss-theor}. A comparison has been drawn whenever
possible between the present approach and the famous Caldeira Leggett
model for the treatment of decoherence and dissipation in quantum
mechanics, showing how the Caldeira Leggett model may
arise as long-wavelength limit of a density-density coupling
preserving translational invariance. This accounts in particular for
the limitation to Gaussian statistics inherent in the Caldeira Leggett
model or variants thereof. At variance with the Caldeira
Leggett model a new microphysical expression for the friction
coefficient has been given, relating it to the Fourier transform of
the interaction potential and a suitable autocorrelation function as
seen in Sect.~\ref{sec:quant-descr-einst}. No need of renormalizations
or energy cutoffs appears in the treatment. Furthermore physical
realizations of the Poisson component of the general structure of
generator of a translation-covariant quantum-dynamical
semigroup~\eqref{eq:34} has been presented, going beyond the typical
restriction to Gaussian statistics.
\par
Even though focusing on the specific issue of Einstein's quantum
Brownian motion, the general results presented in
Sect.~\ref{sec:transl-invar} and~\ref{sec:fluct-diss-theor}, providing
a clearcut connection between expression of the translationally
invariant interaction and precise structure of the associated reduced
Markovian dynamics, fulfilling the natural and physically compelling
requirement of translation covariance, further clarifying the relevant
correlation function of the environment and its connection to the
fluctuation-dissipation theorem, should provide a general framework
for a precise description of dissipation and decoherence in quantum
mechanics, also allowing for a direct connection with microscopic
quantities. Now that experimental quantitative tests of decoherence
begin to be within reach (see for
example~\cite{HarochePRL96-HarocheRMP02,WinelandNature00,PritchardPRL01,ZeilingerQBM-exp}
or~\cite{KieferNEW,Petruccione} for more general references), the next
challenge for the theoretical analysis is in fact no more an
effective, phenomenological description of the phenomenon, but rather
a full-fledged microphysical analysis, in which both phenomena of
dissipation and decoherence can be correctly described.
%\begin{widetext}
%\end{widetext}
%\appendix
\begin{acknowledgments}
   Francesco Petruccione thanks Universit\`a di Milano and Bassano
   Vacchini thanks University of KwaZulu-Natal for kind hospitality.
   Bassano Vacchini is grateful to Prof. L. Lanz for his support
   during the whole work and to Prof. A.  Barchielli for very useful
   discussions on the subject of this paper.  This work was
   financially supported by INFN, and by MURST under Cofinanziamento
   and FIRB.
\end{acknowledgments}

\begin{thebibliography}{99}
\bibitem{citation}
\textit{It is to be hoped that some enquirer may succeed shortly in
  solving the problem suggested here, which is so important in
  connection with the theory of Heat.} (Ref.~\cite{EinsteinDOVER},
p. 18).
\end{thebibliography}

\end{document}